\documentclass[conference]{IEEEtran}
\usepackage{ragged2e,booktabs}
\usepackage[normal]{threeparttable}
\usepackage{float}
\usepackage{cite}
\usepackage{tabulary}
\usepackage{graphicx}
\usepackage{lipsum}
\usepackage{listings}
\usepackage{amsmath}
\usepackage{array}
\usepackage{graphicx}
\usepackage{wrapfig}

\lstdefinestyle{mystyle}{
    basicstyle=\ttfamily\footnotesize,
    frame=single
}
\def\BibTeX{{\rm B\kern-.05em{\sc i\kern-.025em b}\kern-.08em
    T\kern-.1667em\lower.7ex\hbox{E}\kern-.125emX}}

\begin{document}
\title{Reusable Learning Objects: An Agile Approach}
\graphicspath{{./images/}}
\author{\IEEEauthorblockN{R. Pito Salas}
\IEEEauthorblockA{\textit{Computer Science Department} \\
\textit{Brandeis University}\\
Waltham, Massachusetts, USA\\
pitosalas@brandeis.edu}}

\maketitle
\begin{abstract}
This ``Innovative Practice" full paper discusses Reusable Learning Objects (RLOs) and to what extent they have lived up to the promise, particularly of reusability. Reusable Learning Objects have actually been discussed in the literature for the last 20 years and yet true large scale sharing of learning and teaching materials remains relatively rare and challenging. This paper argues that part of the reason is that the granularity of the learning objects that are in use today is not conducive to true reuse. Certainly whole PowerPoint slide decks and word documents are kept in individuals' folders; perhaps bits are cut and pasted and emailed around; but providence, permissions, tracking and tracing are ad-hoc and  styling, formatting, templates, slide layouts need to be tended to repeatedly. It is not an ideal situation. As a result, educators, teachers, course designers, are constantly reinventing the wheel, or searching for where that one excellent assignment, explanation, definition was last seen so it can be copied forward. 

This paper argues that to achieve effective reuse of Learning Objects, the following are required: smaller, more granular (``micro") learning objects; means to combine them into larger presentation products; and modern revision and version control. The paper proposes applying approaches originating in the software engineering community, such as agile methodology, version control and management, markup languages, and agile publishing, which together form the ``Agile Approach" of the title of the paper. With that foundation laid, the paper examines ``CourseGen", an open source software platform designed for creating, sharing, reusing and publishing reusable course content.  CourseGen uses a modified \textit{markdown} format augmented by CourseGen specific directives, such as \textit{\$link\_to} and \textit{\$include\_topic}. The CourseGen compiler converts a collection of CourseGen files into the final format such as a web site or a PowerPoint. CourseGen was designed, used and refined over the last three years in several Computer Science Courses at Brandeis University.
\end{abstract}
\section{Introduction}
This ``Innovative Practice" paper examines ``Reusable Learning Objects"\footnote{In this paper the terms ``learning object", ``content" or ``course content" will be used interchangeably for any part, small or large, structured or unstructured of materials that are presented to learners, as whole slides or web pages, or parts thereof. The word ``teacher" will refer to academics teaching lower, middle and higher education, instructors teaching in a business setting, or anyone in between. The term "course" will refer to a college course, a professional course, a corporate course or program.}(RLOs) in the context of the current literature and practice: what seems to work and doesn't work, and suggests some improvements. It  begins by defining what Learning Objects are and reviews the literature.  This is followed by an examination of the the life-cycle of Learning Objects, from authoring, revising, producing, delivering, and finally reusing. This leads to the introduction of set of factors and features which it is argued will make sharing and reuse easier and more effective. With that foundation laid, the CourseGen framework and tool-set are introduced. CourseGen is particularly focused on creating detailed, flexible and reusable learning objects. The tool-set includes both open source software and a large collection of open source learning objects for Software Engineering and Entrepreneurship courses.
\subsection{What are ``Reusable Learning Objects"?}

Noguera et. al. \cite{Noguera2018AnSystem} define ``Learning Objects"\footnote{From now on, Reusable Learning Object will be abbreviated to RLO and Learning Object to LO} simply: "Learning objects are reusable entities either digital or otherwise used to support learning.", an intuitive but somewhat circular definition. The term has goes back to the '90's and has shown up in the literature from time to time with a variety of analogous definitions. We can understand the phrase ``Reusable Learning Objects" (RLOs) as follows:
\begin{itemize} 

    \item ``Reusable" implies that an existing learning object created for one purpose, could be used (re-used) for another and by another. Satisfying this criterion in turn requires a flexible and appropriate level of granularity; a degree of modularity and composability; a non-proprietary data format; and appropriate permissions from the originator. We will get to each of these in turn.

    \item ``Learning" is easier. An RLO has something to do with learning. Or teaching. Or education. The point is that perhaps the definition should be made broader as there are opportunities for reuse that are not strictly related to learning. Examples might be a grading policy or a teacher-student agreement.

    \item ``Object" probably was used originally to allude to the notion of Objects in the context of Object Oriented Programming, and in that sphere it also has multiple competing definitions. The common thread, and what is meant in the context of RLOs is that an object is small, designed to be reused, and an abstraction that keeps its internal structure hidden.

\end{itemize}

\subsection{Different Types of Reuse}
A challenge that all Educational Technology faces is that teaching styles vary tremendously. This includes the teachers' basic style, their use of media, what is written down ahead of time, during or not at all, use of blackboards, and many others. It is inevitable that approaches that are exquisitely tuned for one teacher are totally unusable by another one. This fact has a profound impact on the efficacy, and generality of any approach. It argues for maximum flexibility in all aspects, but maximum flexibility interferes with reuse. Therefore this paper will not advocate or claim that this approach is the only and best one. Consider these different variations on sharing and reuse
\begin{itemize}
    \item \textbf{Type 1 Sharing with yourself - } It is very common and useful to need an identical or near identical section in more than one course. A fundamental concept, description of a grading policy, a reading list, or biography, may be shared in two courses, or a new version of an old course.
    \item \textbf{Type 2: Sharing with your own department -} less common but still useful is if there are departmentally standard modules from which a teacher might select to construct a new course or a specialized version of a course.
    \item \textbf{Type 3: Sharing with your institution - } It is easy to imagine a repository of standard policies relating to academic honesty, special needs, attendance and so on appear over and over in syllabi and are great candidates for sharing. 
    \item \textbf{Type 4: Sharing with the world - } An open ended availability of building blocks. Basic ideas and concepts are explained by a multitude of courses and teachers. Why reinvent the wheel, if the worlds expert on your topic is willing to share?
\end{itemize}

\section{Literature Review}
This paper focuses on the overall life-cycle, end-to-end, from writing, revision, publication and reuse of ``Reusable Learning Objects."

The concepts of ``Learning Objects" and in particular ``Reusable Learning Objects" are first mentioned (indirectly) in ``An Institutional Web-Based Learning Objects Repository System" by Noguera et. al.\cite{Noguera2018AnSystem}, where they credit Hodgins from Autodesk with the coinage of the term.

Another early paper by Boyle et. al. in `Panning for Gold: Designing Pedagogically-inspired Learning Nuggets" \cite{Boyle2001TOWARDSRE-USE} gives a preview of two of the four types of sharing articulated above, describing how ``nuggets" (learning objects) were shared between different institutions.

Yeassine et. al. have a wonderful chronology and review of the evolution of these ideas in the literature in "Learning Analytics and Learning Objects Repositories: Overview and Future Directions\cite{Yassine2017LearningDirections} which we will not replicate here but is worth looking at.

This paper focuses on specific ways of creating, sharing and reusing Learning Objects such as lecture notes etc. Finding evidence or data for this is more difficult. The bibliography at the end of this paper lists a series of references that while interesting and relevant did not help answer this question.\cite{Lederman2019ProfessorsSurvey}, \cite{Lederman2018SurveyDesigners}, \cite{Klein2019DigitalWeek}, \cite{D2L2019Survey:D2L}

Bartoletti's survey \cite{Bartoletti2013UseAssignments} finally gives some  data. A multiple response survey (respondents can pick more than one) asked what types of Digital Learning Materials (equivalent to our Learning Objects) were used in a recent assignment, PDF, PowerPoint and simple text were the most popular, followed by images, video clips and audio clips. The present paper argues that PDF and PowerPoint content are ``macro" content making them less useful for flexible reuse.

Finally, recent work by Ro et. al ``Org-Coursepack: A Modular and Reusable Teaching Materials Template in Org-mode"\cite{Ro2019Org-Coursepack:Org-mode} comes to many of the same conclusions as the present paper. They recognize the importance of ``file inclusion functionality" in building course materials by assembling and processing textual building blocks. They stress that ``...students are the ultimate beneficiaries of this approach since their overall
learning experience can be enhanced through consistent, properly formatted, strategically presented course materials..."

\section{Where do learning objects come from?}
Teachers invest a lot of time and energy into preparing materials for their courses. There is a wide variation in approaches. These are some of the key distinguishing considerations:

\begin{enumerate}
    \item Timing: Is the course fully "scripted" before it is taught the first time or is there just a general outline which is used to guide day by day decisions.
    \item Medium: Are the course materials delivered as a web site, as a PDF document, as a slide deck, inside an LMS, or some other way?
    \item Structure: Are the materials highly detailed and organized or are they general with important content being revealed only during the delivery, on the blackboard or other medium.
    \item Delivery: Is the course delivered live in a classroom, live online, or asynchronous?
    \item Authorship: Is the course content developed by the teacher, by a course designer, or handed down from previous years?
\end{enumerate}
\subsection{Process}
The process of developing a course follows a certain set of stages irrespective of tools and teacher. What follows is a proposed process framework and description of the stages based on the author's own reading and experience. Not all steps are followed and the order may vary. Depending on the experience and style of the teacher, this process might not really go beyond step one or two before the first day of the course. The purpose of analyzing this process is to create a framework by which we can further analyze challenges at each stage and consider how or whether they can be addressed.

\begin{enumerate}
    \item \textit{Conceptualize the course:} An idea for a course is developed. The barest outlines are formed, for example a sketch of the learning goals, interdependence with other existing courses, a name, etc. 
    \item \textit{Write a syllabus, summary or outline:} Courses generally have a hierarchical structure. At the very least, they are structured into modules or lectures which are delivered in time sequence. Still one can conceive of novel course structures which have modules that can be taken in any order or by any means.
    \item \textit{Collect source materials:} As the course starts coming together, the course creator gets involved in collecting source materials. Papers (literature search), textbooks, articles and web pages are all reviewed for consideration in the course. 
    \item \textit{Discover and retrieve:} In order to be reused, the course materials, the learning objects, have to first be discovered by a teacher. This requires that the available catalog of learning objects need to be organized, tagged, cataloged and easily searchable.
    \item \textit{Write:} This is the meat of the work of course. The course content (in whatever structure, medium or format) has to finally be written down, possibly in detailed narrative, in outline form, PowerPoint slides, diagrams, images or any combination. This is a key point at which reusability comes into play. The granularity, order dependency, and style will determine in fact whether any of these content objects will be reusable. The more they are tied into a specific instance of a specific course, the more difficult it will be to reuse.
    \item \textit{Reorganize:} Natural iteration will lead to reorganization, rewriting, reordering, and cutting to achieve the goals and learning objectives, as well as allotted time, and reasonable breakdown into individual lectures or lessons.
    \item \textit{Find and correct obsolete information} A variety of changes are always required: minor typos, major reorganization, finding information or references that have become dated or obsolete. From time to time more major reorganizations are required. New modules added, ordering changed, and feedback incorporated.
    \item \textit{Publish:} Sometimes, depending on the approach and available tools, the course materials then have to be converted to a new format (e.g. PDF or a web site) and somehow made available to students. This could be through an email, with a CMS, or any of numerous other approaches.
\end{enumerate}

\section{Towards an Agile approach}
It is safe to say that the nirvana of RLOs still lies in the future. While more research is definitely needed, it appears that today most Learning Objects are created, revised and published using single purpose proprietary tools (such as Microsoft Word and Adobe PDF.)

\subsection{Proprietary Tools}
We need to understand the tools and techniques used (and desired) by practitioners in the field. Our preliminary investigations yielded interesting but inconclusive results. We conducted a preliminary, anonymous survey in March of 2020 to the members of the "Instructional Design" Listserve  EDUCASE, a large and active group of educators and instructional designers. The survey' asked them to rank the tool they use to create their learning object, such as PowerPoint, Microsoft Word, Google Documents and many others. 

Here is a summary of the results:
\includegraphics[scale=0.25]{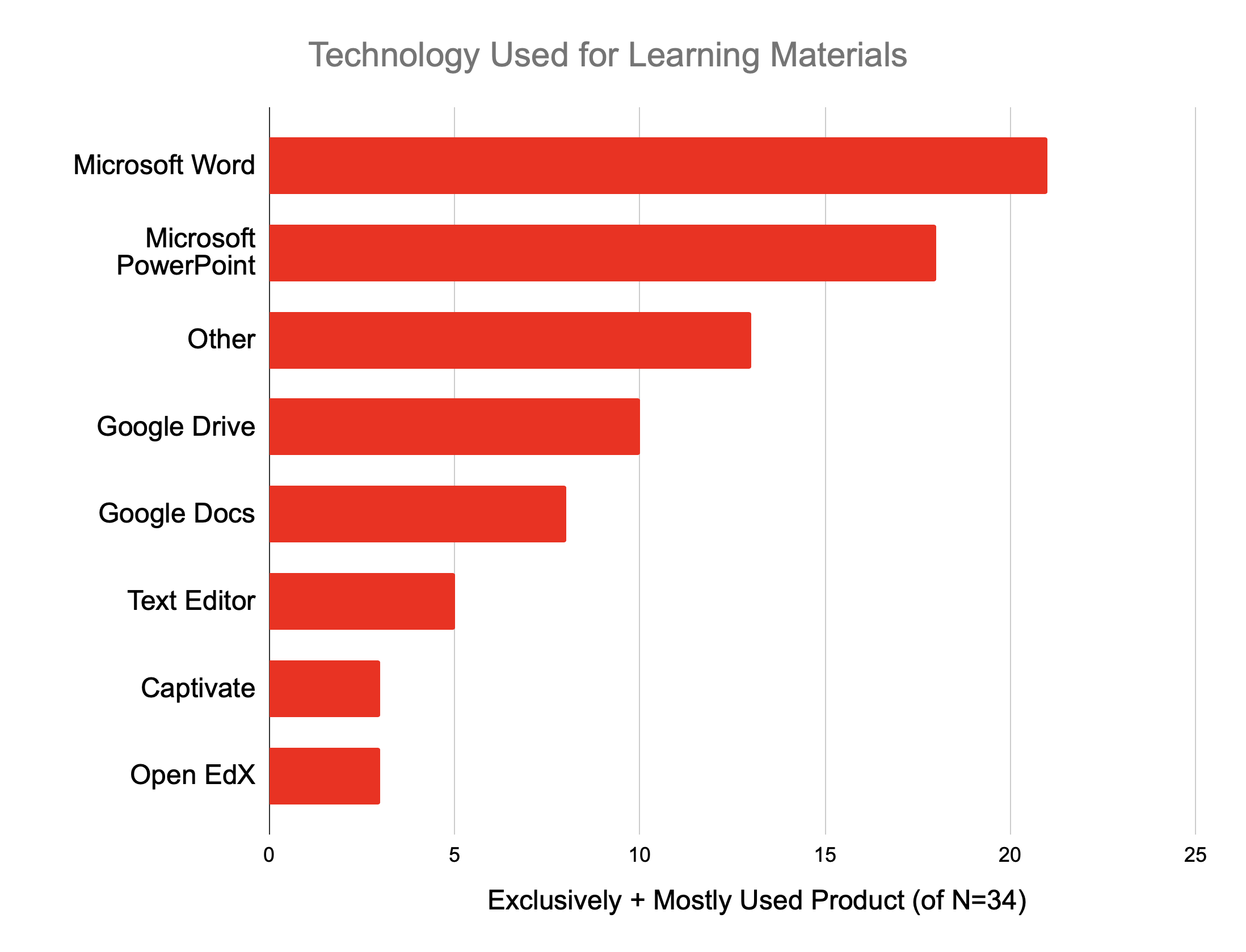}

While there are many advanced products for creation of learning objects, this survey indicates common tools such as Microsoft Word and PowerPoint in the lead. This is not altogether surprising because while there are many other tools they end up diluting the numbers across them. Another study, on the "Use of Digital Learning Materials in Online Course Assignments" \cite{Bartoletti2013UseAssignments} looks at a related but slightly different question and produces a consistent result:

\includegraphics[scale=0.5]{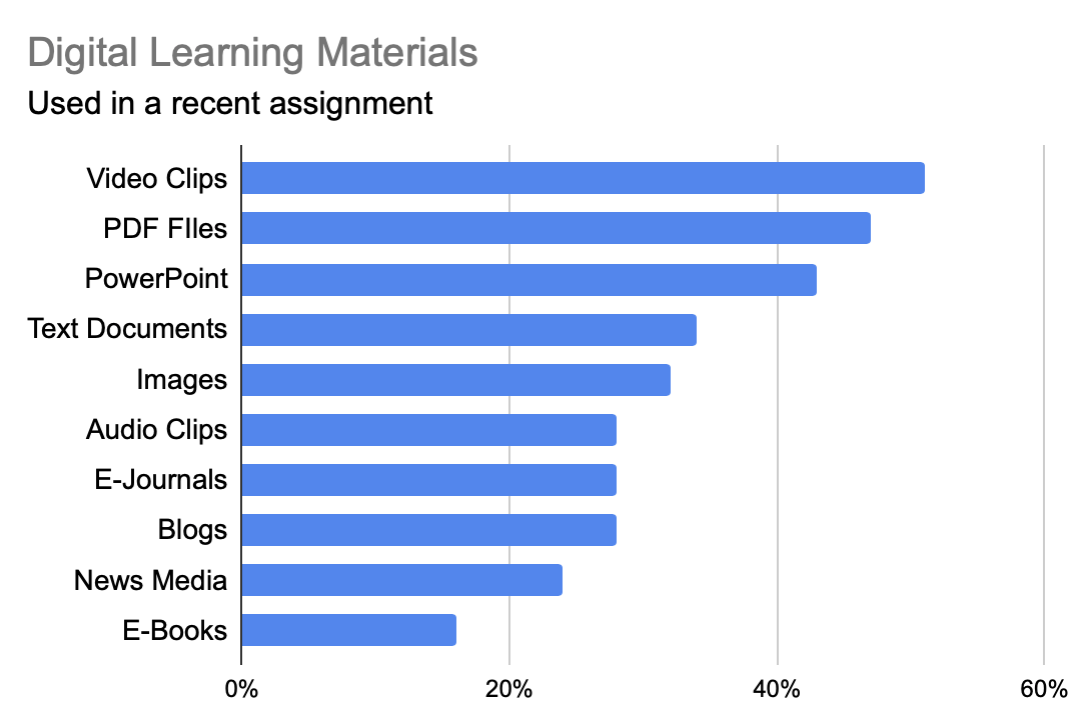}

\subsection{What are the challenges when using these proprietary tools?}
Within this context then, what are the challenges that are encounter in creating and especially reusing our course content? What follows is a high level analysis of these:
\begin{enumerate}
\item \textbf{Cut/Paste Considered Harmful} - Teachers obviously use their previous courses, iterations of courses, and colleagues courses to build and update their own. However this is almost exclusively done by copying (``cut/paste") and duplicating content. The result of this is that when the original bit of content is revised, corrected, or updated, any derived content will become dated. There simply is no way to track and update changes once the content has been copied.

\item \textbf{Revision Management} The process of sharing and revision management is well known and understood in certain domains such as software development. Source code is strictly managed, it is always possible to roll back to a previous version, or to easily see what changes a colleague made. Those and many related capabilities are equally a precondition for RLOs. As long as Learning Objects are copied and pasted, there is no real reuse just duplication over and over again.

\item \textbf{Macro vs. Micro} As will be seen below, reuse and sharing is much more effective when there are truly ``objects": building  blocks, parts of a whole, not complete documents. Learning Objects which are complete PowerPoint presentation decks or multi page PDF documents (``macro"), are far more difficult to share or reuse. Whole documents represent an all-or-nothing scenario which leads to cut/paste/modify, thus defeating reuse. Far richer reuse becomes possible if RLOs are smaller, designed to be modular (``micro") Micro RLOs can be seamlessly assembled from other RLOs into a cohesive result. 

Again as an example, consider the possibility of creating an 80 slide PowerPoint presentation by combining 60 new slides with 20 slides from a repository or library - not by cut/paste but by seamless inclusion references. Correcting an error, changing a date, adding some detail in the repository slide will automatically result in that change appearing wherever it had been included.
\end{enumerate}

\subsection{Agile Approach}
The ``agile approach" described here borrows from "Agile Software Development Process", where it has a long and storied history. The very earliest introduction of the concept was in the "Agile Manifesto" by Kent et. al. \cite{Beck2001AgileManifesto}. 

Since then it has been very well studied. Within Software engineering and process, the Agile approach argues for light-weight processes, adaptive planning, evolutionary development and continual improvement\cite{wikipedia}. Inspired by this, the Agile approach to RLOs has the following characteristics which together enable and encourage Micro RLOs as defined above.

Here are the essential characteristics that we believe are the requirements for a successful approach: 
\begin{itemize}
    \item \textbf{Text files and folders} Use simple, non-proprietary and universal tools to create and manage the Learning Objects (e.g. simple text editors and files)
    \item \textbf{Change management} Use modern change management tools and concepts to manage collaboration and revision histories (e.g. git and GitHub)
    \item \textbf{Easy to read formats} Use a simple, textual format which is easy and efficient to write, edit and read (Learning Object Markup Language, a variant of the Markdown format.)
    \item  \textbf{text files $\rightarrow$ compiler $\rightarrow$ course} Automate the compilation of the text-based Learning Objects into the final presentation format of the course, e.g. PDF or PowerPoint or Web Site or other.
\end{itemize}

\subsection{What about SCORM?}
A comment about SCORM (the ``Sharable Content Object Reference Model"). SCORM is an XML-based standardized description of educational content that may be used to wrap or package content to allow it to more conveniently be stored and shared in compatible CMS systems. SCORM is silent on how the course content is authored, revised or delivered. Therefore while it often comes up in this context, it is fundamentally orthogonal to the present discussion.

\subsection{What about LCMS and LMS?} The topic of this paper is of course intimately related to Learning Management and Learning Content Management Systems (LCMS and LMS), of which there are many examples in the literature -- as open source offerings and as commercial products. Because of the variety of products involved it is not easy to offer clear definition, but generally an LMS is understood to manage course content, and also schedules, assignments, grades, feedback systems and more. So the LMS is a far broader concept that may or may not include a notion of learning objects and even one of reuse of learning objects.

\section{CourseGen}
This paper introduces ``CourseGen", a platform that is built for this paper's ``Agile Approach". It is currently implemented has been in constant use for several years. CourseGen is currently in use by four courses (all by the author) at our institution. We are in discussion with several other instructors to broaden adoption and improve validation of the approach.

The platform is defined in terms of a series of transformations between formatted text files and the final presentation form (for example a web site, set of PowerPoint presentations or PDFs.) We can think of these as a pipeline starting with the content as written by the author, compiled into a format to be published and viewed. Concretely and in our particular implementation, the source format is a variant of the ``Markdown" format and the compiled or output format potentially a web page, a web, site, a PDF document or a PowerPoint document. Each of the outputs are simple transformations of the source content.
lowing parts.
\subsection{Elements}
\subsubsection{CourseGen Learning Objects (CGLOs)}
CourseGen's Learning Objects (CGLOs) are text files created with a generic text editor, using a format known as Markdown modified by the addition of a new set of CourseGen specific directives that follow a fixed and easy to recognize pattern. For example the CourseGen primitive to link to another CGLO is:

\begin{verbatim}$$link_to\end{verbatim}

In addition, each file has a header with one or more properties followed by a body. The set of properties are open ended and the body can be as short or long as required. The decision to require simple text is important as all computer systems can store, edit, email, and export text files: their use sets the stage for all that follows. Here is a small example.
\begin{lstlisting}
---
title: Syllabus
---
#### Course Themes

1. Architecture for Scale: We also want to examine 
how to design systems which will scale under major
load, see $$link_to :scalability_architectures.
\end{lstlisting}
The following is a link to an actual CGLO text file: http://bit.ly/salas01 which compiles into this actual page of course content: http://bit.ly/salas02
\subsubsection{Naming}
To allow course content to be modularized, shared and built up from smaller parts is a simple and intuitive naming scheme is required. CourseGen simply leverages the well understood naming of files and directories, and uses the file path as the name of the CGLO.
\subsubsection{Linking}
Linking is well known in the context of the World Wide Web and Hypertext. A CGLO refers to something outside of itself, and the viewer can click or touch that reference to display the other referenced CGLO. This makes the ``seam" obvious. The viewer is seeing new content in a new context. Both are not viewed at the same time. The benefit of this kind of linking is well established. Within the text of a CGLO, the author can insert a link using the \texttt{\$\$link\_to} primitive.
\subsubsection{Inclusion}
Related but totally different is "inclusion", best explained by a simple example, a CGLO for a Syllabus, called \textit{syllabus}. All syllabi at an institution include a standard paragraph about Academic Integrity. It would be convenient to have a separate CGLO containing the text of this paragraph, called, say \textit{academic\_integrity} as follows:
\begin{verbatim}$$include :academic_integrity
\end{verbatim}

Using the \texttt{\$\$include :academic\_integrity} CGLO directive in the body of the \textit{syllabus} CGLO, the resultant presentation page (be it a web site, a PDF, a PowerPoint) would seamlessly merge in the Academic Integrity policy statement, not as a link, separate slide or page, but literally in place.

\textit{Inclusion} has been overlooked as a key enabler of effective reuse. This primitive makes possible a far more granular reuse of learning objects. By hiding the \textit{seams} between learning objects it becomes possible to reuse one paragraph, assignment description, speaker bio or any of a number of other sub-components to be shared, and then reused seamlessly.

\subsection{Pulling it together}

Using the elements defined above it is possible to now look at the larger objective which is the definition of a full course with a beginning, middle and end. This is done by leveraging the  building blocks introduced above: CGLO files, Naming, Linking and Inclusion semantics with which a richly structured and yet cohesive and seamless course can be structured. 

The course will typically have a clear ``visible" structure (whether it be a web site, a PDF, a PowerPoint or other) built up from a set of CGLOs, some belonging to the course itself and some being pulled from other courses or more typically from a separate and shareable library stored in its own folder structure.

\textit{A library of RLOs} As implied just above, reuse falls out from the Agile Approach. One course can link to or include CGLOs from a library or from another course. Each of those stored simply as a set of files in a tree of directories and sub-directories. Naturally those separate directory trees can be stored on one computer, or far better in a shared and version controlled repository such as GitHub. 
In one shot we leverage the richness of revision control from the world of software engineering in a totally different domain, Reusable Learning Objects. With that come access control, security, revision management, and some more advanced kinds of collaboration such as branching and pull requests. This is a very powerful result of this approach.

\textit{Role of compilation}
To review: Reusable Learning Objects are written using conventional and universal text editors, stored in conventional files and folder, and managed in managed and versioned repositories. 

But certainly we will not ask our learners to read ugly text files. The missing piece is the compilation of course defined as a set of CGLOs into something that is meaningful to the learner. This is accomplished by the CG Compiler. This relatively simple algorithm takes in the complete directory structure, and a series of input parameters and rapidly compiles it into a full web site, or a full series of PDF, or a full series of Power Points or whatever other output formats are contemplated.

\section{Conclusion}
This ``innovative practice" paper introduced an \textit{Agile Approach to Reusable Learning Objects} It explored Learning Objects, and the types of reuse that occur and can occur, looking at some barriers that may still be standing in the way. It defines and describes a framework (``Agile Approach to RLO") for reuse that embodies these principles inspired by well accepted practices from Software Engineering:
\begin{itemize}
    \item Use of simple text files, organized in directories and sub directories as the basic building blocks.
    \item Use of modern revision control techniques to facilitate collaboration and reuse
    \item Use of standard (non-proprietary) tools such as simple text editors and markup languages to allow free sharing
    \item A workflow that ``compiles" the text files into the final delivery platform such as PDF, PowerPoint, a web site, or anything else.
\end{itemize}
This framework is brought to life in the CourseGen platform, software and content library which has been used to define and write, maintain and revise 4 separate courses at our institution. It is very successful in that very narrow context and will be used to further study the validity of the principles described here. 

\subsection{Future Work}
There remains much work to be done. The approach introduced in this paper has a lot of potential as it has only been adopted in a very limited scenario and anecdotally proven its value. However this has to be further studied and measured before conclusions can be reached. Here are some questions for further work. 

\begin{itemize}
    \item There's a need for a new survey of a broader set of teachers to learn how they do their work today
    \item How central are the principles articulated in this paper? What kinds of courses, disciplines, teaching styles are the best fits?
    \item Does the approach articulated here lead to greater teacher productivity and satisfaction?
    \item What kinds of teachers and teaching styles will most benefit from the approach presented here?
    \item How does CourseGen need to evolve to make it suitable for a broader problem space
\end{itemize}

\bibliographystyle{IEEEtran}
\bibliography{references.bib}

\begin{thebibliography}{10}
\providecommand{\url}[1]{#1}
\csname url@samestyle\endcsname
\providecommand{\newblock}{\relax}
\providecommand{\bibinfo}[2]{#2}
\providecommand{\BIBentrySTDinterwordspacing}{\spaceskip=0pt\relax}
\providecommand{\BIBentryALTinterwordstretchfactor}{4}
\providecommand{\BIBentryALTinterwordspacing}{\spaceskip=\fontdimen2\font plus
\BIBentryALTinterwordstretchfactor\fontdimen3\font minus
  \fontdimen4\font\relax}
\providecommand{\BIBforeignlanguage}[2]{{%
\expandafter\ifx\csname l@#1\endcsname\relax
\typeout{** WARNING: IEEEtran.bst: No hyphenation pattern has been}%
\typeout{** loaded for the language `#1'. Using the pattern for}%
\typeout{** the default language instead.}%
\else
\language=\csname l@#1\endcsname
\fi
#2}}
\providecommand{\BIBdecl}{\relax}
\BIBdecl

\bibitem{Noguera2018AnSystem}
\BIBentryALTinterwordspacing
J.~Noguera, S.~Okuboyejo, F.~Ayeni, O.~Sowunmi, and V.~Paindla, ``{An
  Institutional Web-Based Learning Objects Repository System},''
  \emph{International Journal of Current Trends in Engineering {\&} Technology
  www.ijctet.org}, pp. 2395--3152, 2018. [Online]. Available:
  \url{www.ijctet.org,}
\BIBentrySTDinterwordspacing

\bibitem{Boyle2001TOWARDSRE-USE}
T.~Boyle and J.~Cook, ``{Towards a Pedagogically Sound Basis for Learning
  Object Portability and Re-use},'' in \emph{ASCILITE}, 2001.

\bibitem{Yassine2017LearningDirections}
S.~Yassine, S.~Kadry, and M.~A. Sicilia, ``{Learning Analytics and Learning
  Objects Repositories: Overview and Future Directions},'' in \emph{Learning,
  Design, and Technology}.\hskip 1em plus 0.5em minus 0.4em\relax Springer
  International Publishing, 2017, pp. 1--30.

\bibitem{Lederman2019ProfessorsSurvey}
\BIBentryALTinterwordspacing
D.~Lederman, ``{Professors' Slow, Steady Acceptance of Online Learning: A
  Survey},'' 2019. [Online]. Available:
  \url{https://www.insidehighered.com/news/survey/professors-slow-steady-acceptance-online-learning-survey}
\BIBentrySTDinterwordspacing

\bibitem{Lederman2018SurveyDesigners}
\BIBentryALTinterwordspacing
------, ``{Survey of professors shows surprising lack of awareness of
  instructional designers},'' 2018. [Online]. Available:
  \url{https://www.insidehighered.com/digital-learning/article/2018/10/31/survey-professors-shows-surprising-lack-awareness-instructional}
\BIBentrySTDinterwordspacing

\bibitem{Klein2019DigitalWeek}
\BIBentryALTinterwordspacing
A.~Klein, ``{Digital Learning Tools Are Everywhere, But Gauging Effectiveness
  Remains Elusive, Survey Shows - Education Week},'' 2019. [Online]. Available:
  \url{https://www.edweek.org/ew/articles/2019/09/18/digital-learning-tools-are-everywhere-but-gauging.html}
\BIBentrySTDinterwordspacing

\bibitem{D2L2019Survey:D2L}
\BIBentryALTinterwordspacing
{D2L}, ``{Survey: Male Professors Lag Behind Women in Adopting Technology to
  Engage Students | Press Release | D2L},'' 2019. [Online]. Available:
  \url{https://www.d2l.com/newsroom/releases/survey-male-professors-lag-behind-women-in-adopting-technology-to-engage-students/}
\BIBentrySTDinterwordspacing

\bibitem{Bartoletti2013UseAssignments}
R.~Bartoletti, ``{Use of Digiral Learning Materials in Online Course
  Assignments},'' Ph.D. dissertation, Texas Womens University, 2013.

\bibitem{Ro2019Org-Coursepack:Org-mode}
J.~Ro and J.-E. Namkoong, ``{Org-Coursepack: A Modular and Reusable Teaching
  Materials Template in Org-mode},'' \emph{Journal of Open Source Education},
  vol.~2, no.~8, p.~34, 1 2019.

\bibitem{Beck2001AgileManifesto}
\BIBentryALTinterwordspacing
K.~Beck, M.~Beedle, A.~Van~Bennekum, A.~Cockburn, W.~Cunningham, M.~Fowler,
  J.~Grenning, J.~Highsmith, A.~Hunt, R.~Jeffries, J.~Kern, B.~Marick, R.~C.
  Martin, S.~Mellor, K.~Schwaber, J.~Sutherland, and D.~Thomas, ``{Agile
  Manifesto},'' p. 28–35, 2001. [Online]. Available:
  \url{http://agilemanifesto.org/}
\BIBentrySTDinterwordspacing

\bibitem{wikipedia}
\BIBentryALTinterwordspacing
{Wikipedia}, ``{Agile Software Development}.'' [Online]. Available:
  \url{https://en.wikipedia.org/wiki/Agile_software_development}
\BIBentrySTDinterwordspacing

\end{thebibliography}

\end{document}